# How have the Eastern European countries of the former Warsaw Pact developed since 1990? A bibliometric study

Marcin Kozak[&], Lutz Bornmann,* and Loet Leydesdorff #

[&] Department of Quantitative Methods in Economics

Faculty of Economics

University of Information Technology and Management in Rzeszow

Sucharskiego 2, 35-225 Rzeszów, Poland

E-mail: nyggus@gmail.com

*Division for Science and Innovation Studies

Administrative Headquarters of the Max Planck Society

Hofgartenstr. 8

80539 Munich, Germany

E-mail: bornmann@gv.mpg.de

# University of Amsterdam,

Amsterdam School of Communication Research (ASCoR),

Kloveniersburgwal 48,

1012 CX Amsterdam, The Netherlands

E-mail: loet@leydesdorff.net

**Abstract**

Did the demise of the Soviet Union in 1991 influence the scientific performance of the researchers in Eastern European countries? Did this historical event affect international collaboration by researchers from the Eastern European countries with those of Western countries? Did it also change international collaboration among researchers from the Eastern European countries? Trying to answer these questions, this study aims to shed light on international collaboration by researchers from the Eastern European countries (Russia, Ukraine, Belarus, Moldova, Bulgaria, the Czech Republic, Hungary, Poland, Romania and Slovakia). The number of publications and normalized citation impact values are compared for these countries based on InCites (Thomson Reuters), from 1981 up to 2011. The international collaboration by researchers affiliated to institutions in Eastern European countries at the time points of 1990, 2000 and 2011 was studied with the help of Pajek and VOSviewer software, based on data from the Science Citation Index (Thomson Reuters). Our results show that the breakdown of the communist regime did not lead, on average, to a huge improvement in the publication performance of the Eastern European countries and that the increase in international co-authorship relations by the researchers affiliated to institutions in these countries was smaller than expected. Most of the Eastern European countries are still subject to changes and are still awaiting their boost in scientific development.

**Key words**

Normalized citation impact; national comparison; InCites; Warsaw pact; co-authorship

# 1 Introduction

"For decades, the field of science and technology in Bulgaria existed as a system based on immorality, avidity, unscrupulousness, and political maneuvering. Links between science and industry were broken. The main reasons were the feudal structure in science and technology […] and the chaos in the production system." These strikingly dramatic words, written 20 years ago by Denchev (1993, p. 57), describe the situation in the science and technology system in Bulgaria, but one can ascribe them to the science systems in all the Eastern European (EE) countries before the 1990s. The problems listed by Denchev (1993) were among the main reasons for scientific stagnation in the EE countries. The negative influence of the Soviet Union (e.g., with its centrally planned economy and inflexible structures) on scientific developments in its member states as well as its allies in the Warsaw pact was large. Since the breakdown of the communist regimes in the early 1990s, however, the research and development sector in EE countries has had more than 20 years to recover and develop.

The literature about scientific developments in the EE countries provides a pessimistic view. It points out that research in EE countries has had to face many problems, resulting in rather poor scientific performance, especially during the years around the breakdown of the communist regimes (e.g. Egorov 2002; Yegorov 2009). Of course, there were some exceptions in the form of important scientific developments, but they cannot change the general picture of poor scientific performance. Because of the isolation of EE countries from Western countries back in the communist times, researchers from these countries routinely published in journals that were not indexed in the Web of Science (WoS, Thomson Reuters), and accordingly the number of papers published in international journals was rather small. The isolation from institutions and researchers in Western countries resulted in infrequent collaborations with authors in these countries (see also Uzun 2002; Teodorescu and Andrei 2011). After the fall of the communist regime, however, international collaboration networks could have been widened to other (scientifically much stronger) countries.

International co-authorship relations —especially with countries such as the USA— increases visibility of research, thereby improving the citation impact (see Schmoch & Schubert 2008; Glänzel, Schubert & Czerwon 1999; Lancho-Barrantes, Guerrero-Bote & Moya-Anegón 2013). For example, Teodorescu & Andrei (2011) noted an increase in international collaboration after 1989 for Bulgaria, the Czech Republic and Slovakia, Hungary, Poland, and Romania. They also show that for EE countries, internationally co-authored papers are on average cited more than twice as often as those without such collaboration; for Romania this factor is even three times more often. Researchers from EE countries collaborate mainly with Germany and the USA. However, Teodorescu & Andrei (2011) also showed that the demise of the communist regime in various EE countries did not strongly affect international co-authorship relations among researchers from EE countries.

In this study, we analyse the publication development over the last 30 years of all the EE countries. We follow the classification of the United Nations Statistics Division (http://unstats.un.org/unsd/default.htm), which developed groupings of countries and areas, in selecting the EE countries. Among them are those which belonged to the USSR—that is, Russia, Ukraine, Belarus and Moldova—and independent countries that were part of the Warsaw Pact: Bulgaria, the Czech Republic, Hungary, Poland, Romania and Slovakia. Table 1 summarises their respective situations and political status. We shed light on the international collaboration of researchers from the EE countries over time using advanced bibliometric techniques: which developments are visible at three time points (1990, 2000, and 2011) since the breakdown of the communist regimes? Furthermore, we compare the number of publications and normalized citation impact values calculated for the EE countries (listed in Table 1) during the period before, during, and after the breakdown.

The InCites tool (Thomson Reuters) offers a unique opportunity to conduct bibliometric studies of the EE countries because it allows us to (i) use a long publication window (1981 to 2011), (ii) contains categories for differentiation among broad subject areas, and (iii) is suited for the use of statistical procedures in order to obtain an insightful investigation of national citation trends across the years. Using the bibliometric data, we would like to answer the following research questions related to the breakdown of the communist regime:

1. Did the breakdown of the communist regime influence the publication performance (in terms of number of publications as well as citation impact relative to the world average) of researchers in institutions with Eastern European addresses?
2. Did the historical events affect international collaboration by researchers from the Eastern European countries with those of the Western countries?
3. Did these events affect international collaboration among researchers from the Eastern European countries?

## 2 Methods

**2.1 Developments across time using citation and publication data from InCites**

InCites (Thomson Reuters; http://incites.thomsonreuters.com/) is a web-based research evaluation tool allowing assessment of the productivity and citation impact of institutions and countries. The global comparisons module provides citation metrics from WoS for the evaluation of research output of institutions and countries. The metrics are generated from a dataset of several million WoS papers from 1981 to 2011. The metrics for country-specific comparisons are based on address criteria using the whole-counting method: all the addresses attributed to the papers are counted and counts are not weighted by numbers of authors or numbers of addresses.

Country-specific metrics can be downloaded as a national comparison report in Excel format. As a subject area scheme for this study, the main categories for journals of the Organisation for Economic Co-operation and Development (2007) (OECD) were used. InCites provides six further schemes, e.g. the 22 subject areas provided by Thomson Reuters in the Essential Science Indicators. A concordance table between the OECD categories and the WoS subject categories is also provided (at http://incites.isiknowledge.com/common/help/h_field_category_oecd.html). As against the other schemes, the OECD scheme allows for the use of six broad subject areas for WoS data: (1) natural sciences, (2) engineering and technology, (3) medical and health sciences, (4) agricultural sciences, (5) social sciences, and (6) humanities. Each subject area incorporates subordinate fields. Since bibliometric trend analyses requires sufficient publication numbers for each country in each publication year, we considered only the three main subject areas: (i) natural sciences, (ii) engineering and technology, and (iii) medical and health sciences.

Using these three subject areas, the country data (InCites$^{TM}$ Thomson Reuters 2012) were downloaded as an Excel sheet and imported into Stata (StataCorp. 2013) for statistical analysis. According to Marshall & Travis (2011) and Adams (2010), Thomson Reuters calculates the mean citation rate for a country's set of publications and then divides this citation score by the mean of all the publications (in that subject area). A value of 1 for a specific country (in a specific subject area) indicates that the citation impact of papers published by scientists in this country is the same as the worldwide average impact of papers (in this subject area). For example, if the normalized value adds up to 1.2, the corresponding papers were cited on average 20 percentage points above the average (in the subject area). Although the division of means does not provide a proper statistic (Opthof & Leydesdorff 2010; Gingras & Larivière 2011), it can be considered to use these normalized citation impact values at the high level of aggregation of countries (based on several hundreds of publications).

Spearman's rank-order coefficient ($r_s$) for the correlation between publication year and the numbers of publications as well as citation impact are calculated for each country (Bornmann & Leydesdorff 2013; Sheskin 2007). The coefficients support the interpretation of the trend results beyond visual inspection of the curves: a positive coefficient indicates an increasing trend in the country's citation impact or publication output, respectively, across the publication years, while a negative coefficient shows a decreasing trend. To measure the variability of citation impact values across the publication years, the standard deviation (SD) is calculated in the time series for each country and each subject area (see Table 2). This standard deviation indicates the extent of deviations from the mean of a country's citation impacts across all publication years. A relatively small standard deviation, for example, indicates that these impact values do not deviate from the mean across all years to a large extent.

What should be noted here are methodological concerns related to the databases used. InCites provides numbers both for the USSR states separately as well as the USSR altogether (all being included within figures for the USSR) until the beginning of the 1990s. We decided to show the productivity and citation impact numbers for the member states separately, but included the USSR in the co-authorship relations. For this reason, information about number of publications and citation impact relative to world for Russia is visible for 1990 in Figures 1 and 2, but Russia's collaboration network is not visible for 1990 in Figures 3 and 4.

**2.2      Co-authorship relations with authors from other countries**

For the analyses of co-authorship relations, we use the CD-Rom/DVD versions of the Science Citation Index (SCI). This version is not "Expanded" like the WoS version of the Science Citation Index-Expanded (SCI-E), but can be considered as the most policy-relevant set. It includes the most elite and highly-cited refereed journals. The CD-Rom/DVD versions are, for example, used for the Science and Engineering Indicators series of the National

Science Board of the USA (National Science Board, 2012). In 2011, 3,744 journals were included in SCI as against 8,336 journals in SCI-E.

Data for the years 1990, 2000, and 2011 were downloaded and organised in a relational database (see Table 3). Co-authorship relations among countries are counted on the basis of integer counting, but only once. In other words, if a paper contains 3 addresses in country A and 2 in country B, this is considered as a single co-authorship relation (and not as six).

The asymmetrical (2-mode) matrix of documents versus country names is transformed into a symmetrical co-authorship matrix using Pajek v3. A detailed analysis of the global network in 2011 is provided by Leydesdorff, Wagner, Park & Adams (2013); see also Leydesdorff & Wagner (2008) and Wagner & Leydesdorff (2005). The members of the EE countries were assigned (within Pajek) to a first partition; the ten most prolific countries in each year (on the basis of both integer and fractional counting) were assigned to a second partition (e.g. Germany, UK, and the USA). (All other countries worldwide were categorised into a third partition, which was not further considered in this study). Both partitions enable us to draw three maps (for the three years) of the EE nations and three more in relation to ten leading scientific nations at global level. Six map formats are then exported to VOSviewer. VOSviewer maps are based on running mapping and clustering algorithms within VOSviewer (van Eck & Waltman 2010).

In the analyses of the co-authorship relations, Belarus is not present in any year because its co-authorship relations with the other countries were negligible. The German Democratic Republic (GDR) and USSR are only present in 1990, because these names were abandoned after 1991. Although the GDR is not classified by the United Nations Statistics Division as an EE country, we included it into the analyses of the co-authorship relations. The reason is the important position of this country in the co-authorship network of the EE countries.

# 3   Results

Figure 1 shows the numbers of publications and citation impacts calculated relatively to the world averages for the 11 EE countries across the years 1981-2011. Figure 2 shows their normalized citation impact in the three subject areas (natural sciences, engineering and technology, and medical and health sciences). Table 2 summarises the visualised results for the subject areas and countries. In both figures, Spearman's rank-order coefficient for the correlation between publication year and the numbers of publications or citation impacts, respectively, is given for each country.

Two points should be considered in the interpretation of the citation impact results: (1) larger citation impact differences between two following years for one country can be the effect of fewer papers rather than of significant performance differences; and (2) the longer the citation window, the more reliable the performance estimation for a paper is (Research Evaluation and Policy Project 2005). Therefore, the most recent publication years in the figures should be interpreted with care (see, for example, the very high normalized citation impact value for Belarus in Figure 1).

As the results in Figure 1 show, there has been an increasing trend in the number of publications linked to the Czech Republic and Poland, and to a much smaller extent, to Slovakia, Bulgaria, Romania, and Hungary. These countries accessed the EU in 2004 and 2007. For Ukraine, a noticeable decrease in the number of publications was observed in the years 1991-1993; since this time, the number of publications has not yet reached the level of the 1980s.

The citation impact of the EE countries is smaller than the world average (which is represented by the value of 1), both for science in general (Figure 1, Table 2) and for the different subject areas considered (Figure 2, Table 2). However, an increasing trend could be observed for the citation impact relative to the world for all the EE countries (Figure 1), suggesting that in several years at least some of the EE countries can be expected to have a citation impact larger than the world average. Two such very promising countries are the Czech Republic and Hungary, which have already reached the average level for the world.

The country with the largest number of papers, Russia, has low normalized citation impact values, only in recent years approaching the value of 0.5 (that is, two times below the world average); this result refers to science in general as well as the subject areas considered (Figure 2). Moldova illustrates the opposite: a very small country (between 3 and 4 million inhabitants) with a small number of publications (Figure 1), during the last 10 years its normalized citation impact relative to the world increased to the value of around 0.6-0.8. Still below the average, this result is noticeable, especially in the light of the smaller normalized citation impacts of Romania and Belarus, and similar citation impacts of Bulgaria, Poland and Slovakia.

Figures 3a-c show the co-authorship relations among researchers from the EE countries. The darker and the wider the edge, the closer the collaboration is among the researchers from the countries shown. (Hereafter, wherever we refer to "collaboration" between authors, we mean "collaboration that ends up with a common article in a journal indexed in WoS.) The edges are comparable within and between the figures (Figure 4 included).

In 1990, the closest co-authorship relation was between the USSR and Czechoslovakia, and between the USSR and the GDR. There was also quite noticeable collaboration between the USSR and Poland as well as the USSR and Bulgaria. The collaboration of USSR researchers with the non-EE countries was less important than that with colleagues from the other EE countries, which is obvious given the political circumstances in Eastern Europe in 1990 (see above). There was, however, hardly any collaboration between researchers from Hungary and the USSR, or between Hungarian and Romanian researchers.

In 2000, the level of international co-authorship relations by the EE countries had not changed, but its structure had. In Figures 3b-c and 4b-c (so after 1991), the USSR is represented by Russia, Ukraine, and Moldova. As in the USSR, Russian scientists collaborated closely with colleagues from the other EE countries, especially Poland and Ukraine, but also the Czech Republic. There was hardly any collaboration between Slovakian and Russian researchers. In 1990 Hungarian scientists scarcely collaborated with the USSR's, and in 2000 they scarcely collaborated with those in Russia; this changed in 2011, when an increase in this collaboration was observed.

The level of international co-authorship relations among scientists of the EE countries significantly increased in 2011. Russia still made the biggest contribution to this collaboration: its scientists continued to collaborate with Poland, Ukraine, the Czech Republic, and Hungary. But in two other countries, the level of international collaboration with other EE countries increased: Poland (whose researchers also collaborated closely with Russia, the Czech Republic and Ukraine, but also quite closely with Hungary and Romania) and the Czech Republic (mainly with Poland, Russia, Slovakia, but also quite closely with Hungary). Collaboration between researchers from Romania and Hungary, and from Russia and Bulgaria was negligible (and is not represented in the Figures 3c and 4c).

Taking non-EE countries also into account, researchers from most of the EE countries mostly collaborated with authors in the USA (Figure 4a-c), particularly Russia, Poland, and Hungary. Collaboration between the EE researchers and their UK colleagues was much weaker than with those from the USA, which – as the leading scientific country – is the most interesting collaboration partner for all nations. Another important country in terms of scientific collaboration with researchers from the EE countries in 2000 and 2011 was Germany, as could be observed especially for the Czech Republic, Poland, Slovakia, but also for Romania and Bulgaria (Figure 4b-c).

Collaboration among researchers from the EE and Western countries seems to be similar in 1990, 2000 and 2011 (Figures 4a-c). Take Russia: its collaboration with the USA was more efficient (in terms of number of co-authored papers) in 2000 than that between USSR and USA in 1990, but then again, it was less in 2011 than in 2000. Russia's collaboration with the UK in 2000 was similar to that in 2011, but both were slightly stronger than that between the USSR and the UK in 1990. Poland's collaboration with the USA in 2000 and 2011 was slightly stronger than that in 1990, but one might expect a greater difference. These are just two examples, but clearly no boost of international collaboration was observed for any of the present EE countries – rather a small increase or even a lack of any noticeable change.

Although the number of publications by authors from Russia was slightly larger in 2011 than in 2000 and citation impacts were greater in 2011 than in 2000 (Figure 1), collaboration of Russian researchers with those from other countries in 2011 was less intensive than in 2000 (Figures 4b-c). Russia bases its science system on its own infrastructure, know-how and personnel, with only weak co-authorship relations with other countries. This approach results in publication performance which is stable but not increasing: the number of papers with Russian addresses has not changed since 1981, and the normalized citation impact is one of the lowest among the EE countries despite a common opinion that Russian science is very strong.

## 4 Discussion

Our main research question was whether the breakdown of the communist regime affected publication performance in terms of papers published in journals indexed in WoS as well as in international collaboration by researchers from the EE countries with those of the Western countries. Based on the literature one might expect that the answer to this question should be positive. For some of the EE countries (Bulgaria, the Czech Republic and Slovakia,

Hungary, Poland and Romania) an increase in international co-authorship relations was observed after 1989 (Teodorescu & Andrei 2011). The positive answer might also be suggested by common knowledge of the recent history and societal development in the EE countries after the break down of the communist regime.

Take Poland, for example: opening its borders to the Western countries gave rise to quick economic development, but also conditions for scientific development improved. Collaborations with Western colleagues started to be accepted (if not welcomed) by university authorities, universities gained much more autonomy from the government, access to Western literature was acquired, official but illogical and quasi-scientific ideas and paradigms (such as that of lysenkoism; Graham 1974) did not block the development of science, and the like. The picture of Polish science in the last years of the twentieth century was completely different from that during the 1980s.

The situation of former USSR states was much more complex and difficult than that of Poland. When they became independent, they also lost support from the USSR, which actually was the main source of knowledge and funding for science and technology. Fast decentralisation without an appropriate plan can be dangerous, causing the whole scientific system to be crushed. Practically all former Soviet countries suffered from a decrease in funds for research and development, some of which (e.g. Georgia) were dramatic (Yegorov 2009). For these reasons, science development in the former USSR states studied in this paper (Ukraine, Belarus and Moldova) was slower than that in the other EE countries, including a slower development of a network of international collaborators with the Western countries with strong scientific output.

In sum, political opening of borders between the EE and Western countries might indeed lead to an increase in international collaboration among researchers from the EE and Western countries. However, this might be accompanied by decreased international collaboration among the researchers from the EE countries. Our line of thinking was as follows: during the Soviet period, researchers from the EE countries could collaborate with researchers from the other EE countries, and in many situations co-authorship relations with Russia was even expected. Thus, while international collaboration with other (especially Western) countries was hindered and in many cases even impossible, international collaboration among the EE countries in the former Warsaw Pact was possible and sometimes forced.

The drastic change in the political situation which all the EE countries underwent after the breakdown of the communist regime might have given rise to a change in scientific collaboration patterns: what was so difficult and even forbidden before abandoning the communist rule was now allowed (although still not necessarily simple). What is more, one knew that the Western scientists produced more papers, which were more frequently cited, and so many researchers might have been willing to follow the Western standards instead of the Soviet ones. All these facts suggested that international co-authorship relations among researchers from the EE and Western countries should have increased, while among researchers from the EE countries that was not necessarily the case.

From our research it does not necessarily follow that answering the research questions formulated at the end of the introduction section is simple: the situation is more complex and country-specific. The number of papers published by researchers affiliated to institutions located in most of the EE countries did not increase; significant increases were observed for two countries only – the Czech Republic and Poland. For some of the other countries (such as Slovakia and Romania) noticeable changes have been observed only recently (during the recent 5-7 years). Escaping communist rule did not seem to have too great an impact on the scientific performance in these countries.

For Ukraine the situation was even the opposite – after becoming an independent country, Ukrainian researchers published fewer articles in WoS journals, and the number of publications published recently has not yet reached the level of the 1980s. Based on a study of the three Baltic states, namely Estonia, Latvia and Lithuania, Allik (2013) showed that different countries with the same starting positions could develop over the years on completely different trajectories – dependent on policies and decisions made by their policy makers – to end up with different scientific performances. Our results also suggest that a country's scientific system and policy can affect publication productivity. Suffice to look at Russia and Ukraine, which did not develop much over the last 20 years in terms of publication efficiency in journals indexed in WoS even though authors from these two countries published the most papers in such journals in the 1980s.

Close collaboration of the EE countries with the USA obviously results from the USA's scientific prestige, but also from the brain drain in the science sector, which was observed in the late 20th century, as was noted for Ukraine by Egorov (1996, 2010). According to the UNESCO Science Report (2010), of the EE countries Bulgaria, Moldova and Montenegro were still suffering from serious brain drain. Poland – along with India, the UK and Taiwan – is considered one of the four most important countries contributing to a great increase in the number of scientists and engineers in the USA in the 1980s (Sukhatme 1984). Close co-authorship relations with Germany could be explained by its relative proximity to the EE countries (Bornmann, Leydesdorff, Walch-Solimena, & Ettl 2011) as well as the earlier history of Germany: East Germany (GDR) was itself under the USSR's influence.

Collaboration among researchers from the EE countries after the communist regime broke down did not decrease. For most countries the situation was actually the opposite – collaboration increased. Russia is an example; it slightly increased its scientific collaboration with Poland, Hungary, the Czech Republic and Slovakia, and Romania. Of course it is difficult to compare these phenomena in 1990 and in later years, since the Russian network of international collaboration was different from that of the USSR, especially with the inclusion of Ukraine, formerly a Soviet state.

Hungary and Romania are examples of a different phenomenon – their co-authorship relations with the other EE countries increased only recently (which can be observed for 2011 as compared to 1990 and 2000). It suggests that the political and societal changes these countries underwent have had a delayed effect on these countries' scientific systems.

A general picture of international collaboration among researchers from the EE countries is that it seems to have been boosted recently, as can be seen by comparing the situation in 2000 and 2011. Nonetheless, in absolute values international collaboration of EE countries with some of the Western countries, mainly USA and Germany, is still greater than that with the other EE countries. Even, Ukraine's co-authorship relations with the USA are at the same level as that with Russia.

# References

Adams, J. (2010). *Global research report: United Kingdom*. Leeds, UK: Evidence.

Allik, J. (2013). Factors affecting bibliometric indicators of scientific quality. *Trames*, 17(67/62), 3, 199-214.

Bornmann, L., & Leydesdorff, L. (2013). Macro-indicators of citation impacts of six prolific countries: InCites data and the statistical significance of trends. *PLoS ONE,* 8(2), e56768. doi: 10.1371/journal.pone.0056768.

Denchev, S. (1993). Science and technology in the New Bulgaria. *Technology in Science* 15(1), 57-63.

Egorov, I. (2002). Perspectives on the scientific systems of the post-soviet states: a pessimistic view. *Prometheus*, 20(1), 59-73.

Egorov, I. (1996). Trends in transforming R&D potential in Russia and Ukraine in the early 1990s. Science and Public Policy, 23(4), 202-214.

Gingras, Y., & Larivière, V. (2011). There are neither "king" nor "crown" in scientometrics: Comments on a supposed "alternative" method of normalization. *Journal of Informetrics, 5*(1), 226-227.

Glänzel, W., Schubert, A., & Czerwon, H. J. (1999). A bibliometric analysis of international scientific cooperation of the European Union (1985–1995). *Scientometrics*, 45, 185–202.

Graham, L. R. (1974). Science and philosophy in the Soviet Union. New York: Vintage Books.

InCites[TM] Thomson Reuters. (2012). *Report created: 11.02.2013. Data processed: Dec 31, 2011*. Data Source: Web of Science. This data is reproduced under a license from Thomson Reuters.

Lancho-Barrantes, B., Guerrero-Bote, V., & Moya-Anegón, F. (2013). Citation increments between collaborating countries. *Scientometrics*, 94(3), 817-831. doi: 10.1007/s11192-012-0797-3.

Leydesdorff, L., & Wagner, C. S. (2008). International collaboration in science and the formation of a core group. *Journal of Informetrics, 2*(4), 317-325.

Leydesdorff, L., Wagner, C. S., Park, H. W., & Adams, J. (2013). International Collaboration in Science: The Global Map and the Network. *El Profesional de la Información, 22*(1), 87-94.

Marshall, E., & Travis, J. (2011). UK scientific papers rank first in citations. *Science, 334*(6055), 443-443.

National Science Board. (2012). *Science and engineering indicators 2012*. Arlington, VA, USA: National Science Foundation (NSB 12-01).

Organisation for Economic Co-operation and Development. (2007). *Revised field of science and technology (FOS) classification in the Frascati manual*. Paris, France: Working Party of National Experts on Science and Technology Indicators, Organisation for Economic Co-operation and Development (OECD).

Opthof, T., & Leydesdorff, L. (2010). *Caveats* for the journal and field normalizations in the CWTS ("Leiden") evaluations of research performance. *Journal of Informetrics, 4*(3), 423-430.

Research Evaluation and Policy Project. (2005). *Quantitative indicators for research assessment – a literature review (REPP discussion paper 05/1)*. Canberra, Australia: Research Evaluation and Policy Project, Research School of Social Sciences, The Australian National University.

Schmoch, U., & Schubert, T. (2008). Are international co-publications an indicator for quality of scientific research? *Scientometrics*, 74, 361–377.

Sheskin, D. (2007). *Handbook of parametric and nonparametric statistical procedures* (4th ed.). Boca Raton, FL, USA: Chapman & Hall/CRC.

StataCorp. (2013). *Stata statistical software: release 13*. College Station, TX, USA: Stata Corporation.

Sukhatme, S. P. (1994). The real brain drain. Orient Blackswan.

Teodorescu, D., & Andrei, T. (2011). The growth of international collaboration in East European scholarly communities: a bibliometric analysis of journal articles published between 1989 and 2009. *Scientometrics*, 89, 711-722.

UNESCO (2010). UNESCO Science Report 2010: The Current Status of Science Around the World. UNESCO.

Uzun, A. (2002). Library and information science research in developing countries and Eastern European countries: A brief bibliometric perspective. *International Information & Library Review*, 34, 21-33.

van Eck, N. J., & Waltman, L. (2010). Software survey: VOSviewer, a computer program for bibliometric mapping. *Scientometrics*, 84(2), 523-538. doi: 10.1007/s11192-009-0146-3.

Wagner, C. S., & Leydesdorff, L. (2005). Network Structure, Self-Organization and the Growth of International Collaboration in Science. *Research Policy, 34*(10), 1608-1618.

Yegorov, I. (2009). Post-Soviet science: Difficulties in the transformation of the R&D systems in Russia and Ukraine. *Research Policy*, 38, 600-609.

Yegorov, I. (2010). Problems of Measurement of the Real Brain Drain in the Post-Soviet States: the Ukrainian Case. In: Science and Educational Policies in Central and Eastern Europe, Balkans, Caucasus and Baltic Countries. Director of Publication: Engelbert Ruoss Editors: Iulia Nechifor, Elena Severin Collaborators: Gabriela Preda, Rosanna Santesso, Sergiu Porcescu (2010): 141-145.

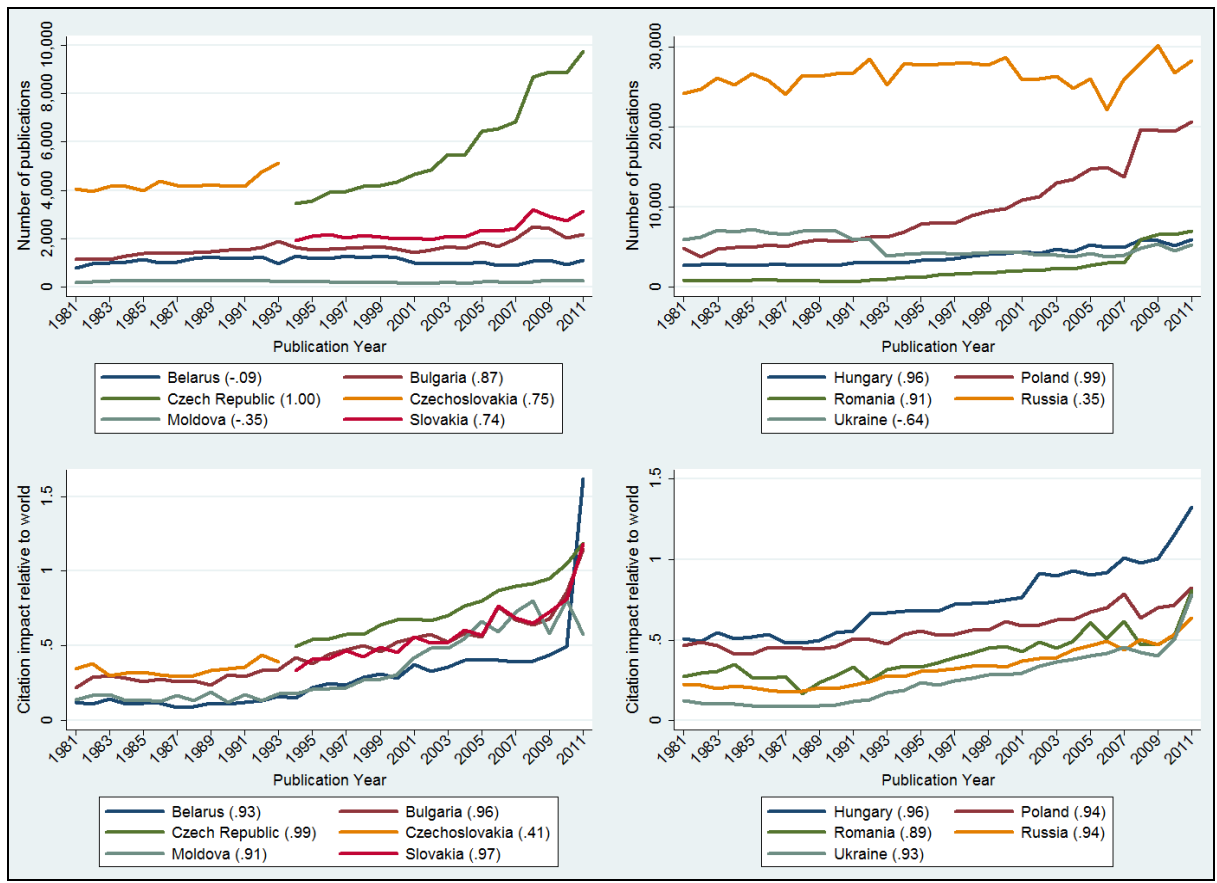

Figure 1. Number of publications and citation impact calculated relatively to the world of 11 Eastern European countries (Belarus, Bulgaria, The Czech Republic, Czechoslovakia, Hungary, Moldova, Poland, Romania, Russia, Slovakia, and Ukraine). In parentheses, Spearman's rank-order coefficient for the correlation between publication year and number of publications/citation impact is given for each country. A high correlation coefficient indicates an increasing or decreasing trend. Source: InCites[TM] Thomson Reuters (2012)

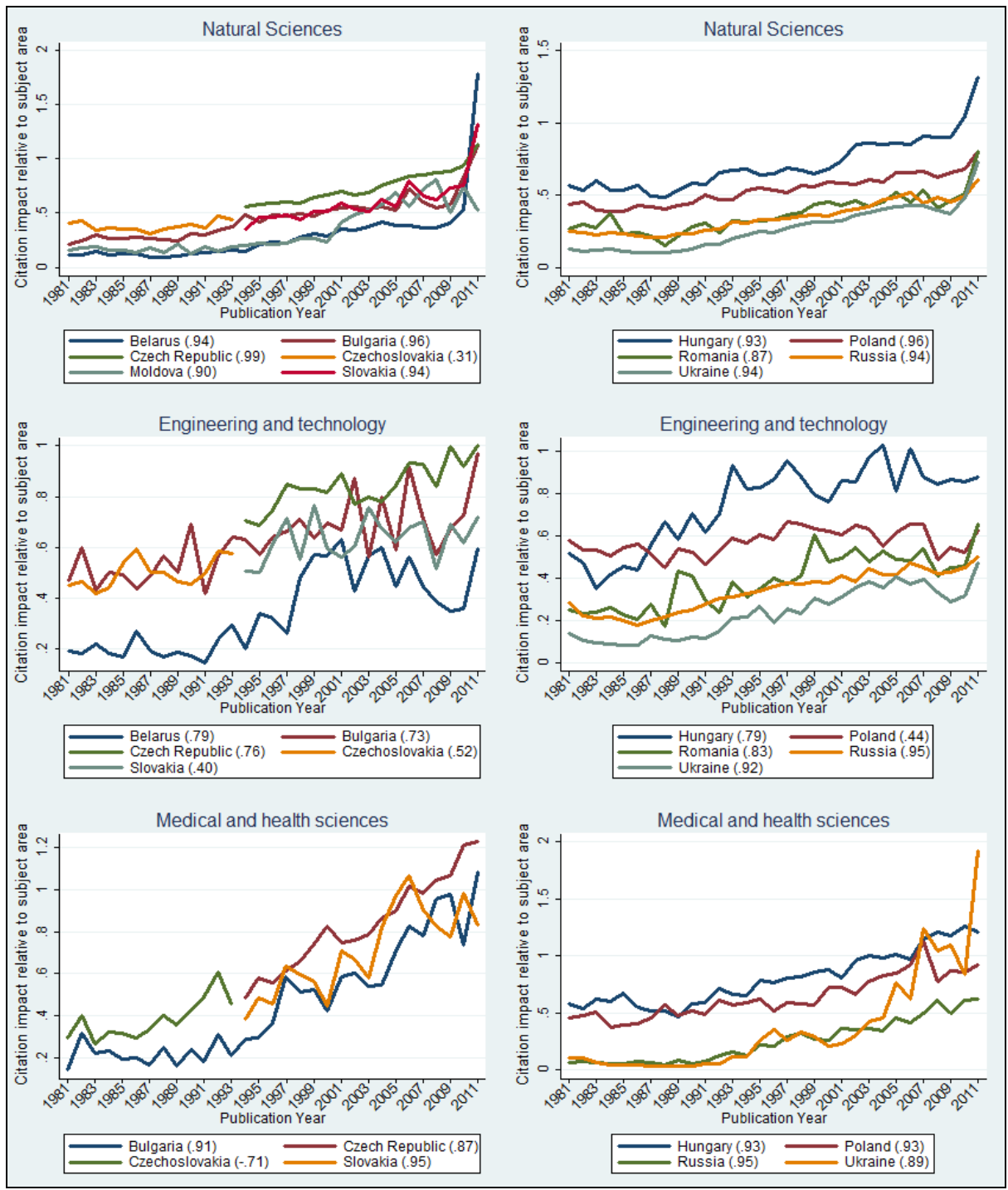

Figure 2. Citation impact of 11 Eastern European countries (Belarus, Bulgaria, The Czech Republic, Czechoslovakia, Hungary, Moldova, Poland, Romania, Russia, Slovakia, and Ukraine) calculated relatively to three subject areas. For each country, Spearman's rank-order coefficient for the correlation between publication year and citation impact is given. A high correlation coefficient indicates an increasing or decreasing trend in citation impact values. Source: InCites[TM] Thomson Reuters (2012). In each subject-specific graph, only countries are considered with at least 100 publications in each publication year.

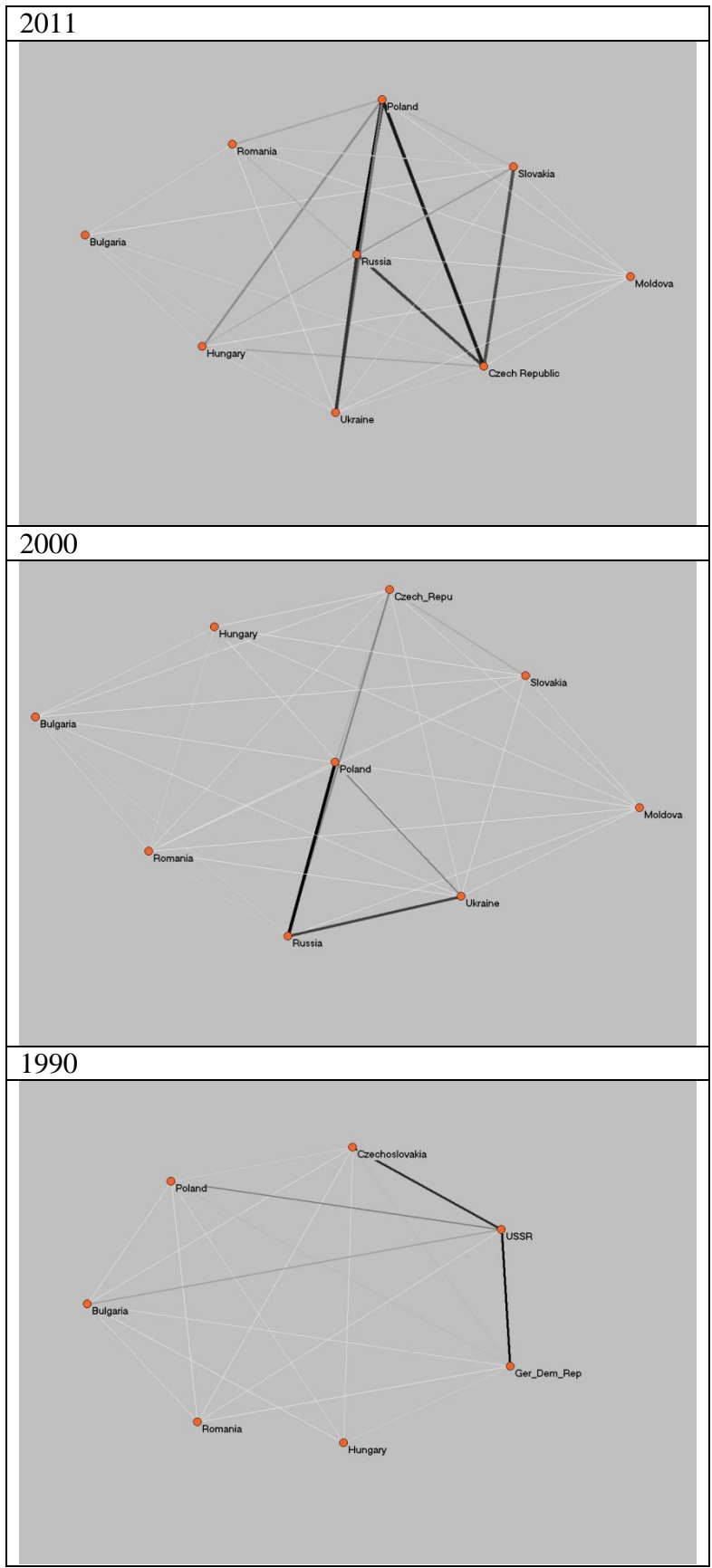

Figure 3. Co-authorship networks of eastern European countries for three time points.

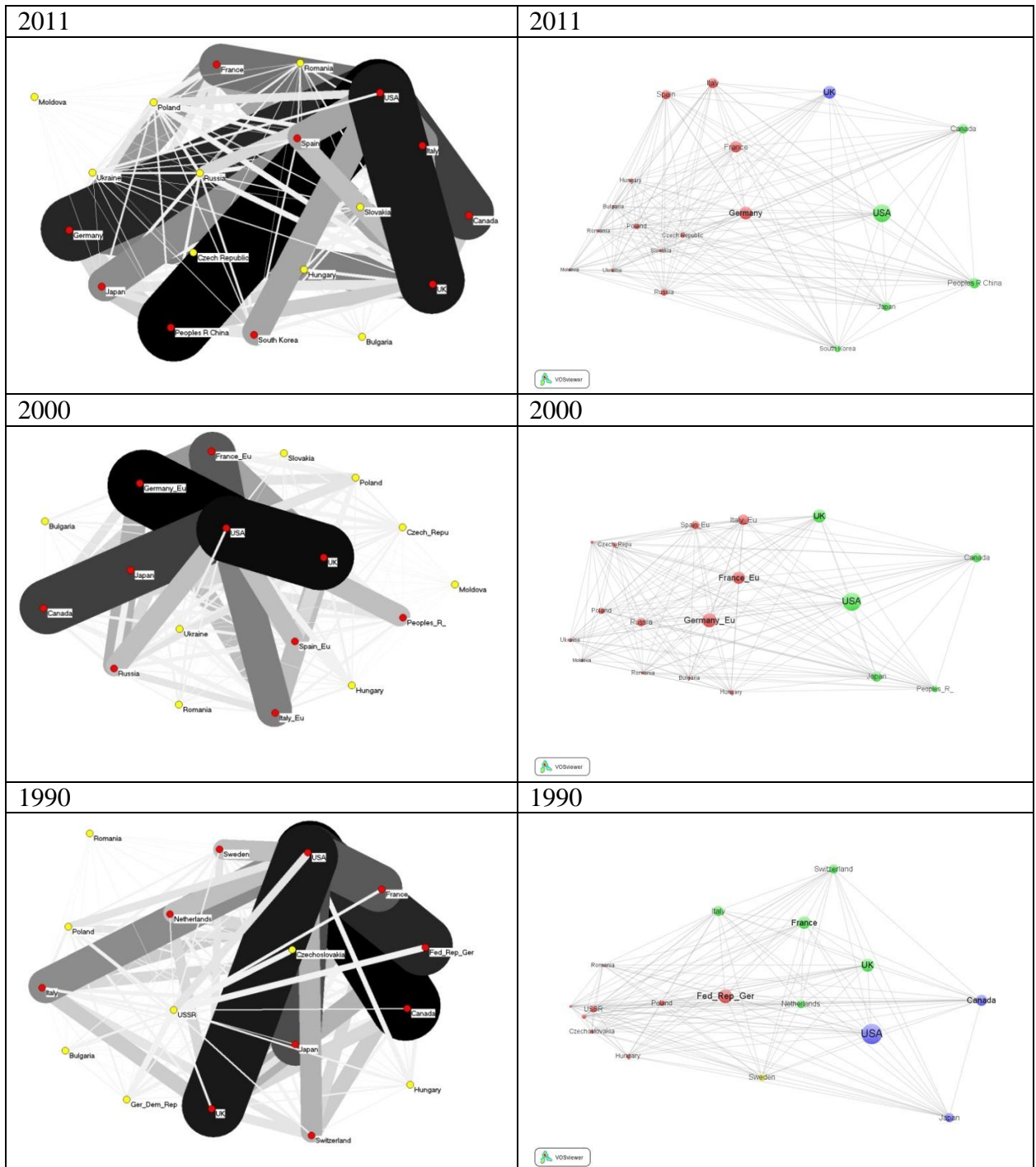

Figure 4. Co-authorship networks of eastern European countries and leading nations at global level for three time points.

Table 1. Summary of current situation in Eastern European countries.

| EE Country | Previous status | Member of EU | Member of NATO | Economic status (according to the World Bank) |
|---|---|---|---|---|
| Belarus | Part of the USSR; sovereign since 1991 | No | No | Developing |
| Moldova | Part of the USSR, sovereign since 1991 | No | No | Developing |
| Ukraine | Part of the USSR, sovereign since 1991 | No | No | Developing |
| Russia | Part of the USSR, since 1991 the Russian Federation | No | No | Developing |
| Bulgaria | Transition and accession since 1989 | Since 2007 | Since 2004 | Developing |
| The Czech Republic | Transition and accession since 1989, independent since 1993* | Yes, since 2004 | Yes, since 1999 | Developed |
| Slovakia | Formerly a part of Czechoslovakia (breakdown of the communist regime in 1989), sovereign since 1993* | Yes, since 2004 | Yes, since 2004 | Developed |
| Hungary | Breakdown of the communist regime in 1989 | Yes, since 2004 | Yes, since 1999 | Developed |
| Poland | Breakdown of the communist regime in 1989 | Yes, since 2004 | Yes, since 1999 | Developing |
| Romania | Breakdown of the communist regime in 1989 | Yes, since 2007 | Yes, since 2004 | Developing |

* Czechoslovakia was split into the Czech Republic and Slovakia in 1993.

Table 2. Minimum, maximum, mean, and standard deviation of citation impact values by subject area and country ($N$ is the number of publication years included in the analysis).

| | **Belarus** | **Bulgaria** | **The Czech Republic** | **Czechoslo-vakia** | **Moldova** | **Slovakia** | **Hungary** | **Poland** | **Romania** | **Russia** | **Ukraine** |
|---|---|---|---|---|---|---|---|---|---|---|---|
| **All subject areas** | | | | | | | | | | | |
| $N$ | 31 | 31 | 18 | 13 | 31 | 18 | 31 | 31 | 31 | 31 | 31 |
| Minimum | 0.09 | 0.22 | 0.50 | 0.30 | 0.12 | 0.34 | 0.48 | 0.41 | 0.17 | 0.18 | 0.08 |
| Maximum | 1.62 | 1.15 | 1.19 | 0.44 | 0.81 | 1.18 | 1.33 | 0.82 | 0.81 | 0.64 | 0.78 |
| Mean | 0.29 | 0.47 | 0.75 | 0.34 | 0.34 | 0.59 | 0.73 | 0.56 | 0.39 | 0.33 | 0.25 |
| Standard deviation | 0.28 | 0.21 | 0.19 | 0.04 | 0.22 | 0.20 | 0.22 | 0.11 | 0.14 | 0.12 | 0.16 |
| **Natural Sciences** | | | | | | | | | | | |
| $N$ | 31 | 31 | 18 | 13 | 31 | 18 | 31 | 31 | 31 | 31 | 31 |
| Minimum | 0.10 | 0.21 | 0.56 | 0.31 | 0.13 | 0.36 | 0.49 | 0.38 | 0.15 | 0.21 | 0.10 |
| Maximum | 1.78 | 1.12 | 1.14 | 0.47 | 0.81 | 1.31 | 1.31 | 0.80 | 0.80 | 0.61 | 0.73 |
| Mean | 0.29 | 0.46 | 0.74 | 0.38 | 0.34 | 0.61 | 0.71 | 0.53 | 0.37 | 0.34 | 0.27 |
| Standard deviation | 0.30 | 0.20 | 0.16 | 0.04 | 0.21 | 0.21 | 0.19 | 0.10 | 0.13 | 0.11 | 0.15 |
| **Engineering and Technology** | | | | | | | | | | | |
| N | 31 | 31 | 18 | 13 | | 18 | 31 | 31 | 31 | 31 | 31 |
| Minimum | 0.15 | 0.42 | 0.69 | 0.42 | | 0.50 | 0.35 | 0.45 | 0.17 | 0.18 | 0.08 |
| Maximum | 0.63 | 0.97 | 1.00 | 0.60 | | 0.77 | 1.03 | 0.67 | 0.66 | 0.50 | 0.47 |
| Mean | 0.35 | 0.63 | 0.84 | 0.50 | | 0.63 | 0.75 | 0.57 | 0.39 | 0.34 | 0.23 |
| Standard deviation | 0.16 | 0.14 | 0.09 | 0.06 | | 0.08 | 0.19 | 0.06 | 0.13 | 0.09 | 0.12 |
| **Medical and Health Sciences** | | | | | | | | | | | |
| N | | 31 | 18 | 13 | | 18 | 31 | 31 | | 31 | 31 |
| Minimum | | 0.14 | 0.48 | 0.27 | | 0.38 | 0.47 | 0.38 | | 0.05 | 0.03 |
| Maximum | | 1.08 | 1.23 | 0.61 | | 1.07 | 1.26 | 1.13 | | 0.63 | 1.92 |
| Mean | | 0.46 | 0.84 | 0.38 | | 0.71 | 0.80 | 0.64 | | 0.25 | 0.37 |
| Standard deviation | | 0.27 | 0.22 | 0.10 | | 0.20 | 0.24 | 0.18 | | 0.19 | 0.45 |

Table 3. Data for co-authorship relations in 1990, 2000, and 2011.

| Data | 1990 | 2000 | 2011 |
|---|---|---|---|
| Articles, reviews, letters (including notes for 1990)* | 508,941 | 623,111 | 787,001 |
| Addresses in the file | 908,783 | 1,432,401 | 2,101,384 |
| Authors | 1,866,821 | 3,060,436 | 4,660,500 |
| Internationally co-authored records | 51,601 | 121,432 | 193,216 |
| Addresses on internationally co-authored records | 147,411 | 398,503 | 825,664 |
| Journals | 3,192 | 3,745 | 3,744 |

Note: * Since 1997, notes are no longer included among citable items in WoS.